\documentstyle[a4,12pt,epsfig,graphicx]{article}
\def\beq{\begin{equation}}  
\def\eeq{\end{equation}}  

\newcommand{\th}{\theta}
\newcommand{\eq}{\begin{equation}}
\newcommand{\eqx}{\end{equation}}
\newcommand{\eqn}{\begin{eqnarray}}
\newcommand{\eqnx}{\end{eqnarray}}

\renewcommand{\th}{\theta}

\newcommand{\rr}[4]{#1, {\it #2 \/}{\bf #3} #4}

\begin{document}
\titlepage
\begin{flushright}
hep-th/0303081 \\
\end{flushright}
\vskip 1cm
\begin{center}
{\large \bf On the AdS/CFT Dual  of Deconstruction 
 }
\end{center}

\vskip 1cm
\begin{center}
{\large Ph. Brax \footnote{Service de Physique Theorique, 
CEA-Saclay F-91191 Gif/Yvette Cedex, France\\
email: brax@spht.saclay.cea.fr}, R.A.
Janik \footnote{Jagellonian University,
Reymonta 4, 30-059 Krakow,
Poland\\
email: ufrjanik@if.uj.edu.pl} and R. Peschanski \footnote{Service de Physique 
Theorique, 
CEA-Saclay F-91191 Gif/Yvette Cedex, France\\ 
email:pesch@spht.saclay.cea.fr}}
\end{center}
\begin{center}
\end{center}
\vskip .2 cm
\begin{center}
\large
\end{center}

\vskip 2cm
\begin{center}
{\bf Abstract}
\end{center}

\vskip .2 cm
\noindent
We consider a class of non-supersymmetric gauge theories obtained
by orbifolding the $N=4$ super-Yang-Mills theories. We focus on the 
resulting
quiver theories in their  deconstructed phase, both at small and large
coupling, where a fifth dimension opens up. 
In particular we investigate the r\^ole played by this extra
dimension  when
evaluating  the rectangular Wilson loops encoding the interaction 
potential 
between
quarks located at different points  in the orbifold. The large coupling 
potential 
of the deconstructed quiver theory is determined using 
the    AdS/CFT correspondence and analysing   the corresponding minimal 
surface 
solution for the dual  gravitational metric. At small  coupling, the 
potential 
between quarks decreases with their angular distance while at strong 
coupling 
we 
find a linear dependence at large  distance along the (deconstructed) 
fifth dimension.

\vskip .3in \baselineskip10pt{}
\bigskip
\vskip 1 cm
\noindent
\newpage
\baselineskip=1.5\baselineskip

\section{Introduction}
Recently there has been a renewed interest in extensions of the
standard model which differ from the supersymmetric framework.
One of the initial motivations for the supersymmetric extensions
of the standard model is the possibility of preserving a
hierarchy between the weak scale and the unification scale.
In supersymmetric theories this springs from the delicate balance between
bosonic and fermionic contributions to radiative corrections.
In particular at the one loop level the quadratic divergences
exactly vanish.
In the context of softly broken supergravity, the supertrace of
the square mass matrix is proportional to the gravitino mass,
hence field independent and preserving the high energy features
of unbroken supersymmetry.
The deconstructed models 
\cite{deconstruct,deconstruct1,deconstruct2,deconstruct3} offer an 
alternative 
to this scenario.
In these models the quadratic divergences are absent due to the 
equivalence
with a fifth dimensional gauge theory forbidding the appearance
of mass terms.

The deconstructed models have a structure highly reminiscent of quiver 
theories \cite{quiver}.
Indeed by considering the field theory limit of $D3$ brane configurations
in the vicinity of an orbifold singularity one can construct 
non-supersymmetric
gauge theories with a product gauge group $U(n)^\Gamma$ and
fields in bifundamental representations. These theories have
proved to be useful in building string realizations of the
standard model \cite{standard model}. Though non-supersymmetric, the 
quiver 
theories
share another feature with deconstructed models. Indeed the
quadratic divergences vanish exactly \cite{brax}. In the deconstructed 
phase of quiver theories, where the gauge group is broken to the diagonal 
gauge 
group
$U(n)^\Gamma\to U(n)_D$, this springs from an underlying
custodial supersymmetry. Similarly to the deconstructed models,
the quiver theories in the deconstructed phase are  equivalent
to a fifth dimensional $U(n)$ gauge theory in the large $\Gamma$ limit.

In this paper we will investigate the properties of (rectangular) Wilson 
loops
for quiver theories in the deconstructed phase. In particular we will
focus on the interaction potential between {\it twisted} quarks, i.e.
quarks corresponding to open strings with end-points in different
sectors of the orbifold cover. In section 2, we recall some ingredients
about quiver theories. In section 3, we compute the quark potential
at weak coupling. The development of a fifth dimension is made explicit
in the $R_5/L^2$ behaviour of the potential. The $L^2$ dependence signals
the propagation of massless degrees of freedom in five dimensions while
the $R_5$ factor is the only dimension-full constant of the five
dimensional theory. In sections 4, 5 and 6 we analyse the strong
coupling behaviour. In section 4, we formulate the computation of the 
rectangular Wilson loop in terms of a minimal surface having the loop as 
boundary (``Wilson surface'') using the 
AdS/CFT
correspondence. In section 5, we analyze the geometry of the  
Wilson surface in and outside the deconstructed 
region. In section 6, we finally compute the potential and thus the force 
between quarks along the deconstructed dimension. We find that the quarks 
have a linear potential at large 
(angular) distance, a property reminiscent of   confinement  along 
the fifth dimension.

\section{Deconstructing Non Supersymmetric Quivers}

We are interested in certain  non-supersymmetric gauge theories
whose structure can be inferred from the world-volume dynamics of D3 
branes
in the neighbourhood of an orbifold singularity. The breaking of 
supersymmetry
is due to the orbifolding which does not preserve the original $N=4$ 
invariance of the low energy dynamics on D3 branes.

Consider the type IIB string theory with a  stack of $n$ coinciding D3 
branes. 
It is well known that the gauge bosons and fermions living on the 
worldvolume 
of 
the D3 branes 
form a 4d $N=4$ supersymmetric Yang-Mills model with  gauge group $U(n)$. 
The six transverse dimensions represent, from the point of view of the 4d 
theory 
living on the 
worldvolume, six extra nongravitational dimensions. The spectrum and 
interactions of that model are the same
as 
the ones obtained by dimensional reduction of the $N=1$ $U(n\Gamma)$ 
gauge 
theory 
living in 10
dimensions.

One can obtain a theory with fewer supersymmetries than 
$N=4$  by dividing the extra dimensions by a discrete group 
$Z_\Gamma$ 
and 
embedding this orbifold group into the gauge group $U(n\Gamma)$.
The spectrum consists of the fields which  are invariant under the 
combined 
geometric and gauge actions of $Z_\Gamma$ (see the quiver diagram in 
Fig.1).

\begin{figure}[ht]
\centerline{
\epsfysize=5cm 
\epsfbox{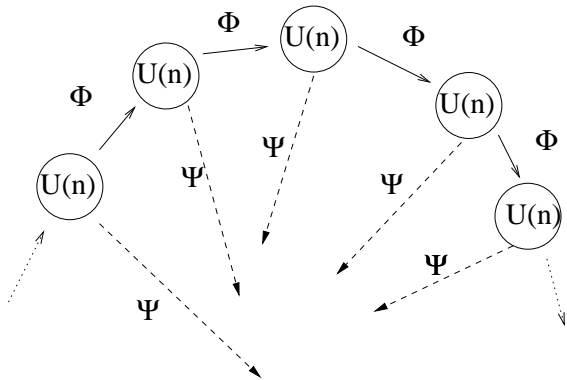}}
\caption{{\it Typical quiver  diagram.} The bosonic $(\Phi)$ and 
fermionic 
$(\Psi)$ fields corresponding to zero-modes of open strings with ends on 
branes 
are schematically represented by arrows. The figure corresponds to  the 
non-supersymmetric case 
$\tilde a_1\equiv 2a_4\equiv 2a_1,$.}
\label{quiver}
\end{figure}

The field  interaction terms  are consistently truncated   to yield 
a smaller daughter gauge theory. The truncation process breaks the gauge 
group 
and 
some (or all) supersymmetries. The gauge symmetry breaking is dictated by 
the 
embedding of the generator of 
$Z_\Gamma$ into $U(n\Gamma)$. The matrix $\gamma$ that represents the 
gauge 
action of $Z_\Gamma$ is chosen to
be of  the form of a direct sum of $\Gamma$ unit matrices of dimensions 
$n 
\times n$,  each multiplied respectively 
by $\omega^i$ with $\omega=e^{ \frac{2 \pi}{\Gamma} i}$. 
The invariant components of the gauge fields
fulfill the condition 
\beq
A = \gamma A \gamma^{-1}
\eeq
where $A$ is a matrix in the adjoint representation  of $U(n\Gamma)$. 
This leaves invariant  the subgroup $U(n)^\Gamma$. 

There are  
four generations of Weyl fermions whose invariant components 
must obey the condition
\beq
\psi^i = \omega^{a_i} \gamma \psi^i \gamma^{-1}
\eeq
where $i=1,..,4$ and 
\begin{equation}
a_1+a_2+ a_3 +a_4 \equiv 0 \ \hbox{mod} (\Gamma)\ .
\end{equation} 
The invariant fermions transform in the 
bifundamental representations  $({\bf n}_l, \bar{\bf n}_{l
+a_i})$ of the broken gauge group
where $l$ numbers blocks of the original $n\Gamma \times n\Gamma $ 
matrices. 
Furthermore, one obtains 
three generations of complex bosons $\phi^i$, $i=1,2,3$, whose invariant 
components fullfil the condition
\beq
\phi^i = \omega^{\tilde{a}_i} \gamma \phi^i \gamma^{-1}. 
\eeq
The invariant scalars transform as $({\bf n}_l, \bar{\bf n}_{l+ 
\tilde{a}_i})$
under the broken gauge group. The integers $\tilde a_i$
correspond to the orbifold action $z_i\to e^{2\pi i \tilde
a_i/\Gamma} z_i$ on the three complex planes. 

The truncated fields have a block structure in the $U(n\Gamma)$ mother 
gauge 
group
\begin{equation}
\phi^i_{lp}=\phi_l^i\delta_{p,l+\tilde a_i},\ 
\psi_{lp}^i=\psi_{l}^i\delta_{p,l+a_i}
\ .
\end{equation}
Supersymmetry is preserved when  the group $Z_{\Gamma}$ is embedded in 
$SU(3)$
\begin{equation}
\tilde a_1+ \tilde a_2 +\tilde a_3\equiv 0 \  \hbox{mod} (\Gamma)
\ .
\end{equation} 
In that case $a_4\equiv 0$ and at least one of the fermions can be paired with 
the 
gauge bosons, i.e. becoming a
gaugino of $N=1$ supersymmetry. 
We focus on the non-supersymmetric case $a_4\ne 0$.

Let us  move a stack of $n$  $D3$ branes from the origin. 
Moving the stacks of $n$ $D3$
branes from the origin corresponds to a diagonal {\it vev} for each 
$\phi^i_l$ 
when $\tilde a_i\ne 0$. This is equivalent to shifting a stack of $n$ 
branes 
from the origin by a distance $R$.
Due to the
$Z_\Gamma$ action, the stacks  have $\Gamma$ copies around the fixed 
point.
The gauge group is broken to the diagonal
subgroup $U(n)_D$. This is the deconstructed phase of the quiver 
theory with a breaking pattern $U(n)^{\Gamma}\to U(n)_D$.

We are interested in the geometry of the orbifold close to the branes.
It is convenient \cite{verlinde,tduality} to parametrize the orbifold with the 
coordinates
\begin{equation}
z_i=r_ie^{i(\alpha_i \theta_1 +\beta_i \theta_2 +\frac{
a_i}{\Gamma}\phi)}
\end{equation}
where the vectors $\alpha_i$ , $\beta_i$ and $a_i$ are orthogonal and
we normalize $\alpha^2=\beta^2=1$.
The flat metric on the orbifold reads
\begin{equation}
ds^2=\sum_{i=1}^{3}dz_id\bar z_i= \sum_{i=1}^3 dr_i^2+ R^2 d\theta_1^2
+R^2 d\theta_2^2 +\frac{a^2R^2}{\Gamma^2}d\phi^2
\end{equation}
where $r_i=R(1+x_i)$ and $x_i<<1$. We have defined $a^2=\tilde 
a_1^2+
\tilde a_2^2 +\tilde a_3^2$ and  assumed that the
orbifold acts non-trivially on the three complex planes.
One recognizes  a circle $S^1$ parametrized by $\phi\in [0,2\pi]$ 
corresponding to the
 $Z_\Gamma$ orbit of radius
\begin{equation}
R_{S^1}=\frac{l_s^2}{R_5}, \ R_5=\frac{\Gamma l_s^2}{aR}\ .
\end{equation}
In the large $\Gamma$ limit, the vicinity of a stack of $n$ branes 
corresponds 
to a cylinder of very small radius $R_{S^1}$.

In this geometry the stack of $D3$ branes become localized at  a point on 
the 
circle $S^1$.
It is identified with its multiple images under the $2\pi$ rotation 
around the 
$S^1$ circle.
It is now easy to analyse the field theory on the stack of $D3$ branes.
Indeed the six dimensions of the orbifold have been replaced by a product
$R^3\times (S_R)^2\times S^1$ where $ S_R$ is a circle of radius $R$. 
Now the field theory of a stack of $n$ $D3$ branes localized in 
$R^3\times 
(S_R)^2\times S^1$
is a $N=4$ $U(n)$ SYM gauge theory corresponding to the massless
open strings joining the stack to itself, see Fig.\ref{quiver}. The 
massive 
states of the theory are 
obtained
by wrapping open strings around $S^1$. The mass spectrum is given by 
\begin{equation}
m_k= \frac{2\pi k}{R_5}
\end{equation}
which is a Kaluza-Klein spectrum of a fifth dimensional theory 
compactified
on a radius
$R_5$.
The appearance of this extra dimension 
can be understood using an appropriate 
$T$-duality \cite{tduality}.
Notice that $R_{S^1}<<l_s$ as soon as $R_5>>l_s$,
i.e.
as $\Gamma \to \infty$ keeping $R$ fixed. In the large $\Gamma$
limit,  substringy
distances
are probed by the D3 branes.  It is more appropriate  to T-dualize along 
$S^1$. 
The radius
of the new circle becomes the large radius $R_5$.
Similarly the $D3$ branes become wrapped $D4$ branes. 
On the $D4$-branes the gauge theory is a $U(n)$ five dimensional gauge 
theory 
compactified on a circle of radius $R_5$. 
This is the generalization to quiver  theories of the deconstructed 
models. 

In the following we will consider the deconstructed quiver theories
both at small and large coupling constant. In particular we
will focus on the Wilson lines.
At weak coupling the Wilson lines can be understood by moving one
of the branes of a stack to infinity and identifying the open
string state connecting the stack of $n$ $D3$ branes to the brane
at infinity as a static quark. Such an open string can wind $w$
times around the orbifold. This is easily pictured on the orbifold
covering space where the string connects $D3$ branes belonging to
different sectors. We will refer to this situation as {\it twisted} 
quarks.

At strong coupling we will
obtain relevant information from the supergravity solution generated by 
the 
stacks of D3 branes. It has to be noticed that the supergravity 
background with 
branes  
displaced to  various points of the circle of radius $R$ has a constant
dilaton. Hence the gauge theory coupling constant does not run with
the energy scale (which is related by the UV/IR principle to the
radial distance scale in the bulk geometry). When we pass to the
orbifold the same is obviously true. 

However one has to keep in mind
that there is a tachyonic twisted mode  at the centre of the 
orbifold which leads to an instability \cite{polch}.  The behaviour of 
the 
dilaton
may become non-trivial  leading to the fact that a running
coupling  may  be generated at this stage.  For the part of the geometry 
relevant for the calculation of the static
potential between static quarks along the deconstructed fifth dimension,
and for $R>> \sqrt {\alpha'}$, the static potential should  not be 
affected by 
this tachyonic instability if the 
change of geometry is confined to a finite region,  close to the centre 
of the 
orbifold \cite{polch}. A more detailed study of this 
instability is out of the scope of the present paper, but certainly 
deserves 
further 
investigation.

\section{Rectangular  Wilson Loop at Small Coupling}

The appearance of a perturbative extra-dimension 
in the large $\Gamma$ regime can be investigated by computing the 
quark-quark potential
in the static approximation \cite{sfetsos}. It is expected that a $1/L^2$ 
dependence of the 
potential in the inter-quark 
distance $L$ should appear as required by  a five-dimensional 
interpretation. 
We will show that this is indeed the case. Moreover we will be interested 
in 
the 
potential between quarks
belonging to different twisted sectors of the orbifold. Indeed in the 
universal 
cover 
of the orbifold one can place quarks in different sectors identified 
under the 
orbifold action. 
We will compute the potential as a function of the angle 
between the quarks in the universal cover. In particular we focus on an 
orbifold 
action on only one plane, i.e.
\begin{equation}
\tilde a_1\equiv 2 a_4, \ \tilde a_2\equiv  0, \ \tilde a_3\equiv 0\ 
\hbox{mod}  
(\Gamma)
\end{equation}
corresponding to
\begin{equation}
a_1\equiv a_4, a_2\equiv -a_4,\ a_3\equiv -a_4\ \hbox{mod} (\Gamma)\ .
\end{equation}
In the following we will focus on the simplest situation $\tilde
a_1\equiv 1\ \hbox{mod}(\Gamma)$,
sketched  in Fig.1.

In this section we will compute the small coupling Wilson loop
for deconstructed quiver theories
\cite{wloops}
\begin{equation}
W=\frac{1}{\Gamma n}\ \hbox{Tr} P e^{ig\int_C (A_m \dot x^m
-X_iu^i\vert \dot x\vert)}
\end{equation}
where $C$ is the Wilson loop contour parametrized by $x^m,$ and
$u^i$ is a unit vector representing the coupling to the six
real scalar fields $X_i,\ i=1\dots 6$ spanning the six extra dimensions.
We will concentrate on the case where the contour is associated to
a rotation along one plane in the six extra-dimensional directions.
The twist of this contour is parametrized by an angle $\Delta
\theta$ representing a rotation between two sectors of the orbifold, see 
Fig.2.

\begin{figure}[ht]
\centerline{
\epsfysize=5cm 
\epsfbox{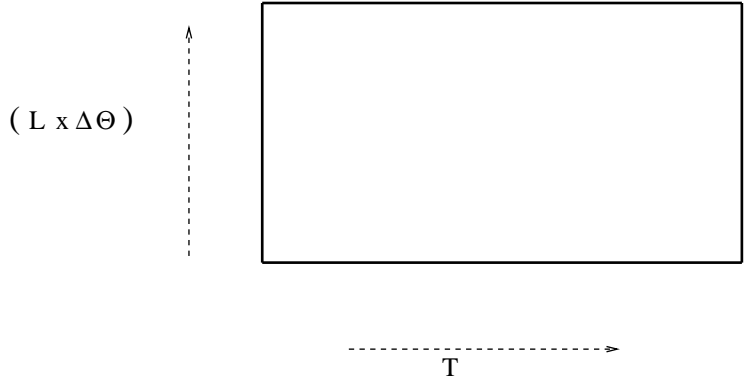}}
\caption{{\it Rectangular  Wilson loop with ``twist''.} The space 
evolution of 
the fields along the Wilson line is accompanied by a corresponding 
rotation 
(``twist'') between sectors of the orbifold, see  text.}
\end{figure}

In our case
\begin{equation}
\vec u=(\cos \Delta \theta \frac{\sigma}{L}, \sin \Delta \theta 
\frac{\sigma}{L})
\end{equation}
in between the two quarks as $\sigma$ runs between $0$ and $L$ while
\begin{equation}
\vec u=(1, 0)
\end{equation}
for the static quark sitting at the origin and 
\begin{equation}
\vec u=(\cos \Delta \theta, \sin \Delta \theta)
\end{equation}
for the static quark at distance $L$. The Wilson loop contour  is 
represented 
in Fig.2.

We also denote by  $T$ the
time length of the Wilson loop. We will take it to be very large 
eventually.
The Wilson loop can be computed by going to Euclidean space and
expanding the connected Green functions to second order in the
gauge coupling constant
\begin{eqnarray}
\ln <W>&=&{g^2 T}\int_C\big(<A_0(0,0)A_0(L,\tau)>+ <\phi
(0,0)\bar\phi(L,\tau)>e^{i\Delta\theta}\nonumber \\
&+& <\phi
(0,\tau)\bar\phi(L,0)>e^{-i\Delta\theta}\ \big)\nonumber \\
\end{eqnarray}
where the first argument of $A_0(0,\tau)$ is at the origin of
space-time where one quark sits and the second is the time $\tau$
parametrizing the time evolution of the quark.
We have denoted $\phi=\frac{X_1+iX_2}{2}$.
In the deconstructed phase one can decompose the fields into
massless and massive modes according to \cite{brax}
\begin{equation}
A_p=\sqrt \frac{2}{\Gamma}\left\{ \sum_{n=0}^{(\Gamma -1)/2} \eta_n
\cos (\frac{2\pi n}{\Gamma}p)A^{(n)}_p +\sum_{n=0}^{(\Gamma -1)/2} 
\sin (\frac{2\pi n}{\Gamma}p)\tilde A^{(n)}_p \right\}
\end{equation}
where each field $A_p$ represents one of the $\Gamma$ blocks and the
index $(n)$ numbers the differents fields. Here we have identified
$\eta_0=1/\sqrt 2$ and $\eta_n=1,\ n\ne 0$.
The fields $A_p^{(n)}$ have masses $m_n$ while the fields $\tilde
A_p^{(n)}$ have masses $m_{\{\Gamma -n\}}.$
Similar expressions hold for the scalar fields with the same mass
spectrum. This fact is due to the underlying custodial
supersymmetry of the non-supersymmetric quiver theories \cite{brax}. 
Using the propagators
\begin{equation}
<A_p^{(n)}(0,0)A_p^{(n)}(L,\tau)>=\int
\frac{d^4p}{(2\pi)^4}\frac{e^{-ip_0\tau +i\vec p \cdot \vec 
L}}{p_0^2+p^2+m_n^2}
\end{equation}
we find that the Wilson loop can be expressed as
\begin{equation}
\ln <W>=g^2n T (1+\cos(\Delta \theta)
\int_0^T d\tau G_5
\end{equation}
where
\begin{equation}
G_5= \sum_{n=0}^{\Gamma -1} \int\frac{d^4p}{(2\pi)^4}\frac{e^{-ip_0\tau 
+i\vec p \cdot \vec L}}{p_0^2+p^2+m_n^2}
\end{equation}
and we  have removed the self energy contributions.
In the deconstruction regime this leads to
\begin{equation}
G_5=R_5 \int \frac{d^5p}{(2\pi)^5}\frac{e^{-ip_0\tau 
+i\vec p \cdot \vec L}}{p_0^2+p^2+p_5^2}
\end{equation}
which is nothing but the five dimensional propagator
\begin{equation}
G_5=\frac{R_5}{ (\tau^2+L^2)^{3/2}}
\end{equation}
For large $T$ we deduce the interaction potential
\begin{equation}
V=\frac{g^2n R_5}{ L^2}\int_0^{\infty}\frac{1+\cos 
\Delta\theta}{(1+u^2)^{3/2}}
=\frac{g^2n R_5}{ L^2}\ \{1+\cos\Delta\theta\}\ .
\end{equation}
Notice the expected $1/ L^2$ behaviour, characteristic of the opening of 
the 
fifth dimension at weak coupling.

The potential is periodic in $\Delta \theta$ in the orbifold cover. 
Antipodal 
points
are such that there is maximal screening with a vanishing
interaction potential. As $\Delta\theta$ increases, the potential
decreases and the interaction force between two twisted sectors 
decreases.
Finally the potential is proportional to the radius of the
compactified fifth dimensional gauge theory $R_5,$ which sets its 
dynamical scale.

\section{The rectangular Wilson Loop at Strong Coupling}

\subsection{Dual  metric deconstruction}
We have seen that in the large $\Gamma$ limit, deconstructed 
quiver theories become similar to a fifth dimensional gauge theories.
We will now analyse the models in the large gauge coupling limit
using the AdS/CFT correspondence \cite{ma98,ma99}.
To simplify the analytic treatment we shall concentrate on the case where 
the
orbifold group acts on a single complex plane. 
The metric due to the presence of the displaced branes is given by
\begin{equation}
ds^2_{10}= H^{-1/2}ds_{4}^2+ H^{1/2}ds_6^2
\end{equation}
where
\begin{equation}
H=1+\sum_{j=1}^{\Gamma }\frac{r_0^4}{\vert r-r_j\vert^4}
\label{H}
\end{equation}
and  $r_j=(Re^{2\pi ij/\Gamma},0,0) $ is the location of the i-th image 
of the 
displaced brane. 
The complement metric $ds^2_6$ is defined up to the orbifold 
identifications.
In the case where the orbifold acts on a single plane we denote by 
$\theta$ the
polar angle in that plane. 
We are interested in computing the Wilson line between quarks 
belonging to different sectors. In the orbifold cover this amounts
to putting quarks in sectors separated by an angle which is a multiple of
$2\pi/\Gamma$. 
At strong coupling the relevant regime is the blown-up vicinity of the 
displaced 
brane where 
\begin{equation}
H \approx \sum_{i=1}^{\Gamma }\frac{r_0^4}{\vert r-r_i\vert^4}
\end{equation}
while the quarks lie at infinity. 

There are various 
relevant regions. Far away from the branes, space-time becomes
isometric to $AdS_5\times S^5/Z_{\Gamma}$. The quarks are on the
boundary of $AdS_5$. In the interior of the six extra dimensions,
the geometry departs from the $AdS_5$ behaviour. In particular
there is a ring around the stacks of branes where the harmonic
function signals the presence of a fifth dimension. The geometrical 
setting in 
the orbifold plane has 
been
sketched in Fig.3.

The strong coupling calculation of Wilson loop expectation values 
reduces, in 
the classical 
approximation of the AdS/CFT correspondence, to the evaluation 
\cite{ma99,Wilson,Gross} of   the bulk  
minimal surface area bounded by the Wilson loop contour. In this context, 
the 
(massive) quarks in the static limit are represented by an open string 
with ends 
near infinity in $r.$

The nature of the minimal surface depends crucially on the
harmonic function $H$ and on the physical length of the wilson
loop $L$. In the following we will consider the region
\begin{equation}
x>> \frac{1}{\Gamma}
\end{equation}
where $x=r/R -1$.
In that region the harmonic function (\ref{H}) can be well approximated 
\cite{sfetsos} by 
\begin{equation}
H=\Gamma r_0^4 \frac{r^2+R^2}{(r^2-R^2)^3}
\label{Hinterp}
\end{equation}
with no angular dependence, i.e. leading to the existence of two
conserved quantities ${\bf E}$ and ${\bf l}$.
Close to the circle containing all the branes, the harmonic
function reads
\begin{equation}
H=\frac{\Gamma r_0^4}{4R^4}\frac{1}{x^3}
\label{Hdecons}
\end{equation}
which is valid for $1/\Gamma << x \le x_*$ where $x_*< {\cal O}(1)$ 
defines 
a  scale  limiting  the  validity of the approximation.
The circle of radius $R^*=R(1+x_*)$ is the outer edge of the
region where the behaviour of the harmonic function is similar to a
five dimensional harmonic function decaying with the third power
of the distance. We will call the region between $R$ and $ R^*$
the  ``deconstruction domain''.

Now, at very large distances the harmonic function becomes the
$AdS_5$ harmonic function
\begin{equation}
H=\frac{\Gamma r_0^4}{r^4}\ .
\label{AdS}
\end{equation}
For  the sake of clarity in  the following calculation and discussion, let 
us denote by $r\equiv \tilde R$ the circle beyond which  (\ref{AdS}) is valid 
with a sufficiently good approximation. This will in particular contain 
the region where the quark sources stand. Between these two regions, 
namely for $R^*<r<\tilde R,$
the harmonic function $H$ of (\ref{Hinterp}) interpolates smoothly.

\subsection{Integrals of motion on the Wilson contour}

\begin{itemize}
\begin{figure}
\centerline{
\epsfysize=10cm 
\epsfbox{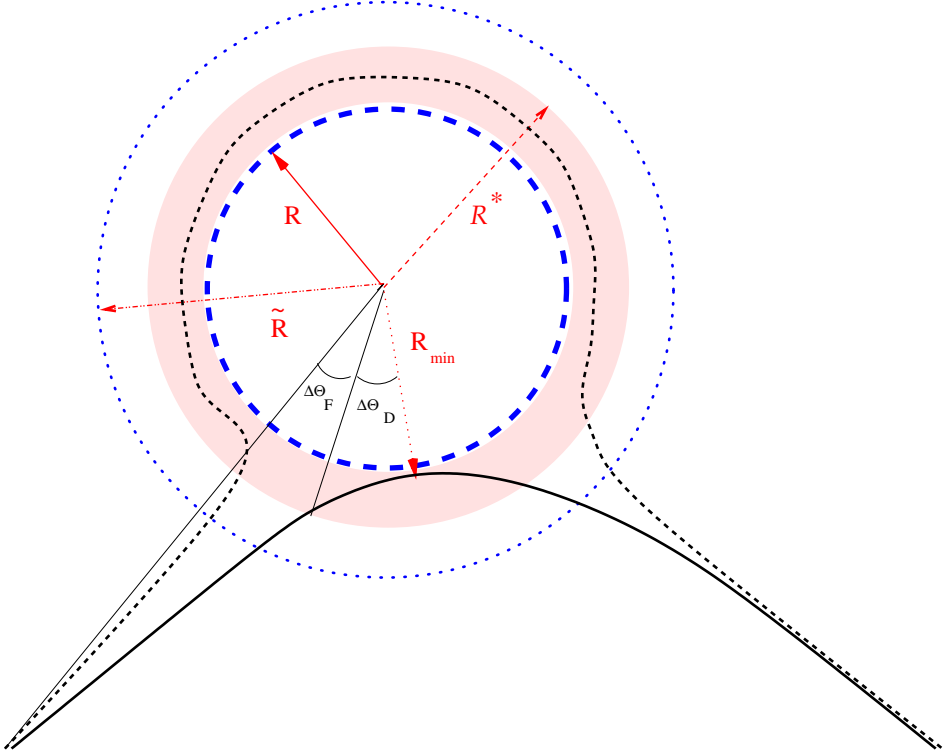}}
\caption{
{\it Wilson surface projected on the ``fifth dimension''.} The minimal 
surface 
contours are displayed  as seen in the orbifold plane. The full ($resp.$ dashed) 
curve correspond to the  dominant ($resp.$ first subdominant) contribution 
to the potential (see subsection 4.2).}
\vspace{-.2cm}
\item $R:$  Location radius of the $D_3$-branes  : {\it Thick  dashed circle}
\vspace{-.2cm}
\item  $R\le r\le R^*:$ Deconstructed phase region  : {\it Hatched ring}
\vspace{-.2cm}
\item $r=\tilde R:$ Lower limit of (approximate ) validity of the  $AdS\times 
S_5$ metrics  : {\it Dotted circle}
 \vspace{-.2cm}
\item $\Delta \Theta_D$ ($resp.\  \Delta \Theta_F$): half the angle 
spanned by 
the 
Wilson contour inside ($resp.$ outside) the deconstruction region (see 
section 
5).
\vspace{-.2cm}
\end{figure}
\end{itemize}

We are interested in computing the interaction potential, {\it i.e.} the 
rectangular Wilson 
loop in the large $T$ limit, 
when
quarks lie in different sectors of the orbifold. 
 Within the AdS/CFT 
correspondence scheme 
the potential is determined by the minimal surface area swept by a string
connecting the boundary quarks \cite{Wilson}. The geometry of this Wilson 
surface, in particular its behaviour in the deconstructed  domain, will 
determine the basic features of the force between quarks.

By comparison with the original evaluations of Wilson contours 
\cite{Wilson}, 
there exist some differences that we have to face. The quark 
sources, represented by  D-branes at  infinity  
are  {\it a priori} outside the deconstruction region, see Fig.3. One 
thus 
has to take care of how and where the minimal Wilson surface is attracted 
near 
the set of orbifolded D-branes. This does require to determine (at least 
with 
sufficient accuracy) the Wilson line contour inside and outside the 
deconstruction region. let us thus first write the general equations 
determining 
the minimal surface area.

This area is determined by the 
surface element
\begin{equation}
dS_2= (H^{-1}+r'^2 +r^2 \theta'^2)^{1/2}d\sigma d\tau\ .
\end{equation}
where $ '\ \equiv d/d\sigma.$
This leads to the potential
\begin{equation}
V(L,\Delta \theta)=\frac{1}{2\pi\alpha'}\int d\sigma (H^{-1}+ r'^2+ r^2  
\theta'^2)^{1/2}\ .
\end{equation}
Let us note that for each fixed $\Delta \th$ we have at least two series 
of 
strings which contribute to the Wilson loop --- one
which stretches `anti--clockwise' with $\theta_f-\theta_i=\Delta \theta + 
2\pi n$,
and one which runs clockwise with $\theta_f-\theta_i= 2n\pi-
\Delta\theta.$ Therefore the 
Wilson loop expectation
value is  given by the infinite sum
\eq
\left\langle W(T\times L) \right\rangle = \sum_{n\ge 0} \ (e^{-T 
V(L,\Delta
\th+2\pi n )} +
e^{-T V(L,2 \pi (n+1)  -\Delta \th)})\ .
\eqx
For  $T\to \infty$ only the two first contributions will survive giving 
the
effective potential
\eq
V_{eff}(L,\Delta \th)= \mbox{\rm min}\left\{ V(L,\Delta\th),
V(L,2\pi-\Delta\th) \right\}
\eqx 
The potential is explicitly periodic. In the following we will
just determine the function $V(L, \Delta\theta)$.

The potential can be cast into the form
\begin{equation}
V=\frac{1}{2\pi\alpha'}\int dr (1+H^{-1}\dot \sigma^2+r^2 \dot 
\theta^2)^{1/2}\ 
,
\end{equation}
where $^{\cdot} \equiv d/dr$.
In the following we shall use the Lagrangian
\begin{equation}
{\cal L}= (1+H^{-1}\dot \sigma^2+r^2 \dot \theta^2)^{1/2}\ .
\end{equation}
where $H$ does not depend on either $\sigma$ or
$\theta$. 
This leads to the existence of two integrals of motion analogous
to the energy and the angular momentum
\begin{eqnarray}
{\bf E}&=&\frac{H^{-1}\dot \sigma }{\cal L}\ =cst.\nonumber \\
{\bf l}&=&\frac{r^2\dot \theta}{\cal L}\ =cst.
\nonumber \\
\label{motion}
\end{eqnarray}
This implies that the Lagrangian can be expressed as
\begin{equation}
{\cal L}= (1-H{\bf E}^2 -\frac{{\bf l}^2}{r^2})^{-1/2}
\label{lagrange}
\end{equation}
The equations of motion can then be deduced
\begin{eqnarray}
\dot \sigma&=& \frac{{\bf E}H}{(1-H{\bf E}^2 -\frac{{\bf 
l}^2}{r^2})^{1/2}}\nonumber 
\\
\dot \theta &=&\frac{{\bf l}}{r^2}\frac{1}{(1-H{\bf E}^2 -\frac{{\bf 
l}^2}{r^2})^{1/2}}
\nonumber \\
\end{eqnarray}
The  total length of the Wilson line will be $2L$ where
\begin{equation}
L= \int_{R_{min}}^\infty \frac{{\bf E}H}{(1-H{\bf E}^2 -\frac{{\bf 
l}^2}{r^2})^{1/2}}dr
\label{L}
\end{equation}
and the twist angle, $2\Delta \theta$, with
\begin{equation}
\Delta \theta= \int_{R_{min}}^{\infty} \frac{{\bf 
l}}{r^2}\frac{1}{(1-H{\bf E}^2 
-\frac{{\bf 
l}^2}{r^2})^{1/2}}dr\ .
\end{equation}

It is important to note that the minimal surface will not reach the 
circle of 
branes
but stops at a minimal value $R_{min}$ where the Lagrangian
${\cal L},$ see (\ref{lagrange}) goes to infinity.  More precisely, since 
then $
{dr}/{d\sigma} \equiv 1/\dot \sigma \propto {\cal L},$ the minimized 
Wilson 
contour will follow a trajectory near $R_{min},$ spanning an angle 
$2\Delta\Theta_D$ inside the deconstruction region, and 
$2\Delta\Theta_F$ outside this region, see Fig.3.

\vspace {1cm}

\section{Dual  geometry of deconstruction}

\subsection{Length and twist of the Wilson line in the deconstruction domain}
Let us first focus on the contribution to the length $L_D$ of the Wilson 
line 
in  the
deconstruction region.

Considering (\ref{L}), together with the harmonic function 
(\ref{Hdecons}), and 
using an 
appropriate rescaling of variable, one obtains
\begin{equation}
L_D=\frac{1}{x_{min}^{1/2}}\frac{\sqrt \Gamma 
r_0^2}{R}\int_1^{x_{*}/x_{min}}\frac{du}{u^{3/2}\sqrt{u^3-1}}
\end{equation}
where we require that $r=R(1+x)$ lies within the deconstruction
region. One gets 
\begin{equation}
x_{min}=\frac{\Gamma 
r_0^4}{L_D^2R^2}F^2\left(\frac{x_{*}}{x_{min}}\right)
\end{equation}
where
\begin{equation}
F(y)=\int_1^{y} \frac{du}{u^{3/2}\sqrt{u^3-1}}\equiv \frac 23 
(y^3\!-\!1)^{\frac 
12}\ y^{\frac 72}\ _2F_1(7/6,1;3/2|1\!-\!y^{-3})
\ .
\end{equation}
Looking for a solution where $x_{min}/x_*<<1$ for $R_{min}$ to lie in the
deconstruction region, we find to leading order 
\begin{equation}
x_{min}=\frac{\Gamma r_0^4}{L_D^2R^2}F^2
\end{equation}
where 
\begin{equation}
F\equiv F(\infty)= \frac{2\sqrt \pi\Gamma(\frac{2}{3})}{\Gamma 
(\frac{1}{6})}
\approx .86237\ .
\end{equation}
Imposing that $1/\Gamma << x_{min} << 1$ implies that the deconstruction 
regime 
is characterized by
\begin{equation}
 \frac{F (g^2n)^{1/2}R_5}{\sqrt \Gamma}<<L_D<<F (g^2n)^{1/2}R_5
\ .
\label{limit}
\end{equation}
Note that (\ref{limit})
has the meaning of a limitation of the 4-dimensional 
distance $L_D$ over which the potential calculation can be done within 
our 
approximation scheme. Larger 4D distances will be briefly mentioned
in the conclusion.

We find that
the  integrals of motion (\ref{motion}) can be expressed as
\begin{eqnarray}
{\bf E}&=& \frac{2R^2}{\sqrt \Gamma r_0^2}
x_{min}^{5/2}\frac{1}{(x_{min}^2+ 
\frac{\Delta\theta_D^2}{G^2})^{1/2}}\nonumber 
\\
{\bf l}&=& 
\frac{\Delta\theta_D}{G}\frac{R}{(x_{min}^2+\frac{\Delta\theta_D^2}{G^2})
^{1/2}
}
\ .
\label{integrals}
\end{eqnarray}
Hence the twist in the deconstruction region, which we choose as the 
starting 
value for our 
evaluation can be identified with
\begin{equation}
\Delta \theta_D= \int_{x_{min}}^{x_*} \dot \theta \ dx\ .
\end{equation}
This leads to
\begin{eqnarray}
{\bf E}&=& \frac{2R^2}{\sqrt \Gamma r_0^2}
x_{min}^{3/2}\frac{1}{(1+ \frac{\Delta\theta_D^2}{x_*^2})^{1/2}}\nonumber 
\\
{\bf l}&=& 
\frac{\Delta\theta_D}{x_*}\frac{R}{(1+\frac{\Delta\theta_D^2}{x_*^2})^{1/
2}}
\ .
\label{integralbis}
\end{eqnarray}
since
\begin{equation}
G=\int_1^{x_{*}/x_{min}}du \frac{u^{3/2}}{\sqrt{u^3-1}}
\ ,
\end{equation} 
and $G\sim x_*/x_{min}$ diverges linearly as $x_{min}/ x_*<<1$.

Using the previous results {\it cf.} (\ref{Hdecons},\ref{integralbis}), 
we find 
that
\begin{equation}
H {\bf E}^2\equiv \frac {\Gamma r_0^4\ {\bf E}^2 }{4R^4 x^3} 
<\frac{1}{1+ 
\frac{\Delta 
\theta_D^2}{x_*^2}}\equiv \frac {{\bf l}^2 }{R^2} 
\left(\frac{\Delta\theta_D^2}{x_*^2}\right) 
^{-2}\ ,
\label{inequality}
\end{equation}
therefore ${\bf l}^2/r^2$ is much larger than $H{\bf E}^2$ for
$r\ge R^*$ provided
\begin{equation}
\frac{\Delta \theta_D}{x_*}>> 1.
\label{condition}
\end{equation}
Given $\Delta \theta_D,$ which will ultimately define the  quark separation 
distance 
along the 
deconstructed dimension, we thus choose the cut-off $x_*$ in such a way
as to 
verify the 
condition (\ref{condition}).
This implies that the width of the deconstructed region we consider is 
sizeably 
smaller than the angle covered by the Wilson line in this
deconstruction region.

\subsection{Length and twist of the Wilson line outside the deconstruction 
domain}

Since the quarks sources are initially placed at infinity, one cannot make 
definite conclusions on the interquark potential without studying the outside of 
the deconstruction domain. Let us thus discuss the geometrical features of the  
Wilson surface solution when 
 $r>R^*.$ Following the indications of Fig.3, 
the discussion may  imply two regions, one with $r>\tilde R$ where one can use 
the conformal 
$AdS_5 \times S_5$ metrics and one transition region $R^*<r<\tilde R.$

The discussion depends mainly on the range of values of the twist $\Delta 
\theta_D$ one considers within the deconstruction domain. If we are considering 
$\Delta \theta_D$ large enough, which will correspond to a large distance in the 
deconstructed fifth dimension, then the relation (\ref{condition}) stands for 
finite values of $x_*.$ This means that the transition region $R^* <R< \tilde R$ 
is small and thus  not so much contributing to the potential.

Using then (\ref{inequality}), one considers valid the condition 
\begin{equation}
H{\bf E}^2 << \frac{l^2}{r^2}\ .
\end{equation}
This implies that
the Lagrangian for $r >  R_*$ becomes 
\begin{equation}
{\cal L}=\frac{r}{(r^2-l^2)^{1/2}}\ .
\label{Lagrange}
\end{equation}
This is notably different from the case of non-rotating strings in the 
$AdS_5\times S_5$ case,
where the $H\bf E^2$ term is the only non-trivial contribution  in the
Lagrangian.
Notice that the previous approximation is valid in all the region
outside the deconstructed domain.
The contribution to the string length  outside the
deconstructed region is
\begin{equation}
L_F= 2\ \frac{{\bf E}\Gamma r_0^4}{{\bf l}^3}\int_{R^*/{\bf l}}^\infty 
\frac{u^2+u_R^2}{(u^2-u_R^2)^3}
\frac{u }{(u^2-1)^{1/2}}du 
\end{equation}
where $u_R=R/l$. 
This is negligible for $x_{min}<<1$ as it scales as $ x_{min}^3$
while $L_D$ scales like $1 /\sqrt{x_{min}}$. 
We then obtain that
\begin{equation}
L\approx L_D
, L_F << L_D\ ,
\end{equation}
an approximation which  improves for large $L$.

The situation is different for the bending of the string (see Fig.3).
The bending angle of the string in the far-away region is given by
\begin{equation}
\label{e.thetaf}
\Delta \theta_F=
\int_{r^*/{\bf l}}^{\infty}\frac{du}{u(u^2-1)^{1/2}}\equiv \pi/2-\arctan 
\sqrt{\left(\frac {r^*}{\bf l}\right)^2-1}
\end{equation}
which is a finite quantity.
For large values $\Delta\theta_D/ x_*>>1$ we find that ${\bf l}=R$ and $
\Delta\theta_F \to \frac {\pi}2 .$
Notice that the bending in the far-away region saturates. Hence it is 
possible 
to ascribe a 
definite initial position to the source quarks in the far-away region in 
such a 
way that the minimal Wilson contour 
travels over the deconstruction region for a given value of 
$\Delta\theta_D.$

If we are now considering $\Delta\theta_D$ smaller, which will correspond to  
smaller distances in the deconstructed fifth dimension, then the relation 
(\ref{condition}) requires   $x_* \ll 1\ .$ This also means that the transition 
region $R^* <R< \tilde R$ is large and will require using the interpolating 
function $H$ of (\ref{Hinterp}) for the minimization. Without considering  this more 
complex minimization problem, let us note that the function (\ref{Hinterp}) 
contains singular tems in $1/x^2$ and $1/x$ which are expected to modify the 
interquark potential obtained from the $1/x^3$ term of (\ref{Hdecons}). All in 
all, this means that our predictions are only valid for the large angular 
distance behaviour of the potential.

\section{The static interquark potential: results and comments}

We can now be more specific and evaluate the various
contributions to the static potential for the large distance regime (large 
$\Delta \theta_D$).

The potential from the deconstruction region is simply
\begin{equation}
V_D=\frac{Rx_{min}}{2\pi\alpha'}\sqrt{G^2 x_{min}^2+\Delta\theta_D^2}
\ .
\end{equation}
As $L_D$ increases  this gives
\begin{equation}
 V_D= \frac{ n g^2 F^2 R_5}{2\pi
L_D^2}\sqrt{x_{*}^2+\Delta \theta_D^2}\ .
\end{equation}
The first term leads to the expected $1/L^2$ behaviour for a
five-dimensional theory. When $\Delta \theta_D>> x_*$ this reads
\begin{equation}
 V_D= \frac{ n g^2 F^2 x_5}{2\pi
L_D^2}
\end{equation}
where we have put  $x_5=R_5 \Delta\theta_D$.
The   potential is proportional to the 't
Hooft coupling $g^2n$ while the potential is in
$(g^2n)^{1/2}$ for the usual AdS case.
Moreover $V_D$
is both proportional to $R_5$ and $\Delta\theta_D$. It is tempting
to interpret the combination $x_5=R_5 \Delta \theta_D$ as a fifth
dimensional distance between the quarks. In that case, the
behaviour of the potential is reminiscent of confinement along
the fifth dimension. It has to be noted that the 
would-be string tension
\begin{equation} 
{\cal T}= \frac{g^2 n  F^2 }{2\pi
L^2}
\end{equation}
depends on the four-dimensional distance between the quarks, but this dependence 
is valid only in the limited region $L << (g^2n)^{1/2} R_5$ ({\it cf.} 
(\ref{limit})).

Let us briefly comment on the physical meaning of the potential
$V_D(\Delta\theta_D,L_D)$. If we had inserted  the probe D3
branes (carrying the Wilson loops) at the edge of the deconstructed
region,  i.e. at $R^*$, we would have had  massive $W$-- bosons in the theory and 
$V_D(\Delta\theta_D,L_D)$ would be  related to their interaction
potential. However in order to make the sources infinitely massive,
and hence to have the analogues of quarks in the fundamental
representation, we have to move the probe D3 branes and the
Wilson loop away to $r \to \infty$. Then the $\theta$ parameter is 
modified by $\theta_F$ i.e. $\Delta\theta \to
\Delta\theta_D+\Delta\theta_F$ according to (\ref{e.thetaf}). This
drives the angles $\Delta\theta$ to larger values and changes the potential. 
Nevertheless the contribution to the potential from the far away
region is nearly trivial as is shown by the following computation. 
Hence, the main  contribution to the potential comes from the
deconstructed region and can be interpreted as the interaction
potential as seen by static quarks on the edge of the deconstructed domain.

For completeness, 
let us estimate the contribution to the potential from the domain 
far outside 
the deconstruction region. One obtains
\begin{equation}
\Delta V=V_F-V_0
\end{equation}
where $V_F$ is the contribution for $r\ge R^*$ and $V_0$ is a
regularization associated with an infinitely massive straight
string. 
The contribution from the far-away region is 
\begin{equation}
V_F-V_0= \frac{l}{2\pi\alpha'}\int_{r_*/{\bf l}}^\infty du [ 
\frac{u}{(u^2-1)^{1/2}}-1]\equiv \frac{1}{2\pi\alpha'}\left({r^*}-\sqrt{ 
{r^*}^2-{\bf l}^2}\right)
\end{equation}
Notice that this contribution is independent of $L$. We are
particularly interested in the large $\Delta\theta_D>> x_*$ region.
In that case ${\bf l}=R$ and therefore this reduces to a constant
contribution to the potential.

Let us end with some comments on our results. Having analysed non supersymmetric 
quiver theories in their
deconstructed phase both at small and large coupling, we have
retrieved  the expected fifth dimensional behaviour.
In particular,  examining the angular dependence of the quark potential
we have found a striking difference between the small and large
coupling regimes. In the former, the force between the quarks decreases
and vanishes for antipodal quarks. This implies that
static quarks behave like non-interacting particles as long as
they sit at antipodal points in the orbifold. 
At strong coupling we have found that the potential increases linearly
with the angular distance, mimicking a confining behaviour along the
large compact extra dimension.

These results are valid as long as $L<<(g^2n)^{1/2} R_5$. In
the deep infrared regime where $L>>(g^2n)^{1/2} R_5$, the open
string probes a region where $x<<1/\Gamma$. In this regime the
harmonic function $H$ becomes identical with the one of a single
stack of $n$ D3 branes. This corresponds to the fact that at very
low energy the deconstructed quiver theory is a $N=4$ $U(n)$
gauge theory. As well-known from deconstructed models, it is only
in a finite energy range, i.e. a finite interval in $L$, that the
gauge theory looks five-dimensional.

The situation would be entirely different if we had considered
$R\le \sqrt{\alpha'}$. In that case, the non-supersymmetry of the 
configuration
is signaled by the presence of a twisted tachyon deforming the
geometry of space-time. The analysis of the dual gauge theory in
this case is beyond the scope of this paper\cite{polch}.

\subsubsection*{Acknowledgements}

R.P.  wishes to thank Stefan Pokorski for a fruitful discussion.
RJ would like 
to thank the Service de Physique Th\'eorique for hospitality when part of 
this 
work was
 done. RJ was supported in part by KBN grant~2P03B09622.


\begin{thebibliography}{99}

\bibitem{deconstruct} \rr{Nima Arkani-Hamed, Andrew G. Cohen, Howard 
Georgi}{Phys.Rev.Lett.}{86}{(2001) 4757};\\ 
\rr{Christopher T. Hill, Stefan Pokorski, Jing 
Wang}{Phys.Rev.}{D64}{(2001) 
105005}.

\bibitem{deconstruct1}\rr{Csaba Csaki, Graham D. Kribs, John 
Terning}{Phys.Rev.}{D65}{(2002) 015004};\\ 
\rr{Hsin-Chia Cheng, Konstantin T. Matchev, Jing 
Wang}{Phys.Lett.}{B521}{(2001) 
308};\\
\rr{Nima Arkani-Hamed, Andrew G. Cohen, Howard 
Georgi}{}{}{hep-th/0108089};\\ 
\rr{Piotr H. Chankowski, Adam Falkowski, Stefan Pokorski}{JHEP 
0208}{}{(2002) 
003}



\bibitem{deconstruct2}\rr{ C. Csaki, J. Erlich, C. Grojean, G. 
Kribs}{Phys.Rev.}{D65}{(2002) 015003};\\ 
\rr{Tatsuo Kobayashi, Nobuhito Maru, Koichi 
Yoshioka}{}{}{hep-ph/0110117};\\
\rr{Nima Arkani-Hamed, Andrew G. Cohen, Howard Georgi}{JHEP 
0207}{}{(2002) 
020};\\ 
\rr{Z. Chacko, E.Katz, E.Perazzi}{}{}{hep-ph/0203080}.



\bibitem{deconstruct3}
\rr{Nima Arkani-Hamed, Andrew G. Cohen, Howard 
Georgi}{Phys.Lett.}{B513}{(2001) 
232};\\
\rr{H-C. Cheng, C. Hill, J. Wang}{Phys.Rev.}{D64}{(2001) 095003};\\ 
\rr{N. Arkani-Hamed, A. G. Cohen, T. Gregoire,J. G. Wacker}{JHEP 
0208}{}{(2001) 
4757}; \\ \rr{Tatsuo Kobayashi, Nobuhito Maru, Koichi 
Yoshioka}{}{}{(2002) 020}.



\bibitem{quiver} \rr{Michael R. Douglas, Gregory Moore}{D-branes, 
Quivers, and 
ALE Instantons}{}{hep-th/9603167};\\
\rr{Michael R. Douglas, Brian R. Greene, David R. Morrison}{ 
Nucl.Phys.}{B506}{(1997) 84};\\
\rr{S. Kachru, E. Silverstein}{Phys.Rev.Lett.}{80}{(1998) 4855}.


\bibitem{standard model} \rr{G. Aldazabal, S. Franco, L. E. Ibanez, R. 
Rabadan, 
A. M. Uranga}{JHEP 0102}{}{(2001) 047}.




\bibitem{brax} \rr{Philippe Brax, Adam Falkowski, Zygmunt Lalak, Stefan 
Pokorski}{Phys.Lett.}{B538}{(2002) 426};\\
\rr{Philippe Brax, Zygmunt Lalak}{Acta Phys. Pol.}{B33}{(2002) 2399}.


\bibitem{verlinde} \rr{Sunil Mukhi, Mukund Rangamani, Erik Verlinde}{JHEP 
0205}{}{(2002) 023}. 

\bibitem{tduality} \rr{N. Arkani-Hamed, A.G. Cohen, D.B. Kaplan, A. 
Karch, L. 
Motl}{Deconstructing (2,0) and Little String Theories}{}{hep-th/0110146}. 


\bibitem{polch} \rr{A. Adams, J. Polchinski, E. Silverstein}{JHEP 
0110}{}{(2001) 029}. 



\bibitem{sfetsos} \rr{Konstadinos Sfetsos}{Nucl.Phys.}{B612}{(2001) 191}. 
 

\bibitem{wloops} \rr{A.A. Tseytlin, K. Zarembo}{Wilson loops in N=4 SYM 
theory: 
rotation in S5}{}{hep-th/0207241}. 


\bibitem{ma98} \rr{J. Maldacena}{Adv. Theor. Math. Phys.}{2}{(1998)
231};\\ 
\rr{S.S. Gubser, I.R. Klebanov and
A.M. Polyakov}{Phys. Lett.}{B506}{(1998) 105};\\
\rr{E. Witten}{Adv. Theor. Math. Phys.}{2}{(1998) 253}.

\bibitem{ma99} \rr{O. Aharony, S.S. Gubser, J. Maldacena, H. Ooguri
and Y. Oz}{Large $N$ field theories, String Theory and
Gravity,}{}{hep-th/9905111}.

\bibitem{Wilson} \rr{J. Maldacena}{Phys. Rev. Lett.}{80}{(1998) 4859};\\
\rr{S.-J. Rey and J. Yee}{Macroscopic strings as heavy quarks in large
$N$ gauge theory and anti-de Sitter supergravity,}{}{hep-th/9803001}.

\bibitem{Gross} \rr{D.J. Gross and H. Ooguri}{Phys. Rev.}{D58}{(1998) 
106002}.

\bibitem{witten}\rr{E. Witten}{Adv. Theor. Math. Phys.}{2}{(1998) 505}.


\end{thebibliography}
\end{document}